\documentclass[letter]{IEEEtran}       

\usepackage{setspace}
\usepackage{placeins}
\usepackage{afterpage}
\usepackage[utf8]{inputenc}
\usepackage[cmex10]{amsmath}
\usepackage{graphicx}
\usepackage{amssymb}
\usepackage{amsthm}
\usepackage{subfigure}
\usepackage{soul}
\usepackage{cite}
\usepackage{color}
\usepackage{caption}
\usepackage{amsmath}
\usepackage{setspace}
\usepackage{tabu}
\usepackage{array}
\usepackage{booktabs}
\usepackage{amssymb}
\usepackage{threeparttable}
\usepackage{float}
\usepackage{stfloats}
\DeclareMathSizes{10}{9}{7}{6}
\IEEEaftertitletext{\vspace{-3\baselineskip}}

\begin{document}

\title{Extended Kalman Filter Beam Tracking for Millimeter Wave Vehicular Communications}

\IEEEoverridecommandlockouts
\author{
    \IEEEauthorblockN{Sina Shaham, Matthew Kokshoorn, Ming Ding, Zihuai Lin, and Mahyar Shirvanimoghaddam}\\
    \IEEEauthorblockA{
        School of Electrical and Information Engineering, The University of Sydney, Australia \\
        Email: \{sina.shaham, matthew.kokshoorn, zihuai.lin, mahyar.shm\}@sydney.edu.au, ming.ding@data61.csiro.au}
}

\maketitle
\begin{abstract}

Millimeter-wave (mmWave) communication is a promising technology to meet the ever-growing data traffic of vehicular communications. Unfortunately, more frequent channel estimations are required in this spectrum due to the narrow beams employed to compensate for the high path loss. Hence, the development of highly efficient beam tracking algorithms is essential to enable the technology, particularly for fast-changing environments in vehicular communications. In this paper, we propose an innovative scheme for beam tracking based on the Extended Kalman Filter (EKF), improving the mean square error performance by $49\%$ in vehicular settings. We propose to use the position, velocity, and channel coefficient as state variables of the EKF algorithm and show that such an approach results in improved beam tracking with low computational complexity by taking the kinematic characteristics of the system into account. We also explicitly derive the closed-from expressions for the Jacobian matrix of the EKF algorithm.  

\end{abstract}


\section{Introduction}

Advanced technologies such as high definition road maps, real-time updates, and a large number of sensors embedded in vehicles can result in enhanced safety of passengers, parking assistance for drivers, and better vehicular traffic management systems~\cite{fazliu2019mmwave, kawser2019perspective}. The common ground for enabling all such revolutionary technologies in vehicular communications is achieving a higher network capacity for data sharing. 

One of the most appealing candidates to improve network capacity is exploiting the Millimetre Wave (mmWave) spectrum~\cite{ghosh2019inclusive,r2}. In fact, the use of mmWave is not a new concept for applications such as vehicle to vehicle (V2V) and vehicle to infrastructure (V2I) communication \cite{r3}. However, it was only until recently that directional antenna systems have made the technology a practical option by concentrating the power on one or more beams. Such enhanced designs have alleviated many of the existing issues, such as high signal attenuation and blockage probability of communication in the mmWave spectrum. 

Despite numerous advantages of directional antenna arrays, they result in narrow beams that require frequent channel estimation, particularly in fast-changing environments such as vehicular communications. Therefore, there is a need for more advanced algorithms that improve beam tracking in order to reduce the number of channel estimations required. In this paper, our goal is to adopt the Extended Kalman Filter (EKF) for beam tracking in the mmWave vehicular communications by proposing an innovative scheme and providing closed-form expressions.



Some of the prior works are briefly introduced as follows. The authors in \cite{midc} proposed a beamforming protocol for the 60-GHz propagation channel. The method exploited training sequences for the detection of signal strengths. The evaluation of the proposed algorithm was provided in \cite{midc2}. In contrast to our work, the approach required multiple beam training sequences that impose significant overhead on the system. The most relevant schemes to our approach are the works in \cite{zhang} and \cite{rh}. In \cite{zhang}, a full scan of all possible beam directions was proposed in order to achieve the measurement matrix. The proposed scheme applied the EKF algorithm to track the arrived beams; However, it required a high overhead of pilot transmission for this purpose. The state model was based on the angle of arrival (AoA) and angle of departure (AoD), and no consideration was given to the channel coefficient. In \cite{rh}, the authors improved the tracking performance by having a single measurement instead of a full scan. The system model was also based on AoA and AoD with the addition of the channel coefficient.

To apply the EKF algorithm for beam tracking in vehicular communications, the system is modeled based on the so-called state evaluation model and observation expression. There are several reasons why current approaches based on EKF are not suitable for mmWave vehicular communications. First, the state evolution model is considered to be linear, which is not the case in a vehicular communication setting. Second, current approaches such as~\cite{zhang,rh} assume that the noise in the system is additive, which is also not valid for vehicular communications as it will be demonstrated in this paper. Third, based on the current scheme, it is not possible to take the impact of kinematic characteristics of vehicles into consideration. For example, no provisions can be made for characteristics such as the initial speed, noise on the speed, and the duration of transmission blocks.


In this paper, we propose a new scheme to enable beam tracking based on the EKF algorithm for mmWave vehicular communications. We explain our approach using a simple V2I scenario. As opposed to the previous approaches that the state variables were based on AoA and AoD, we propose to use position, velocity, and the channel coefficient as state variables. Based on the proposed approach, we are able to significantly reduce the complexity in the calculation of Jacobians required for modeling the system. We explicitly derive the expressions needed for the Jacobians, and finally, implement the EKF algorithm on a V2I scenario. Our proposed scheme improves the beam tracking performance in the mmWave spectrum, considers real-world characteristics of vehicular communications, such as velocity and the transmission block duration, and reduces the computation complexity when calculating the Jacobians. In particular, the proposed scheme improves the mean square error of AoA and AoD by $49\%$.\\

\textit{Notation} : We use capital bold-face letter ($\boldsymbol{A}$) to denote a matrix, $\boldsymbol{a}$ to denote a vector, $diag(\boldsymbol{a})$ to denote a diagonal matrix whose diagonal entries are $\boldsymbol{a}$ starting in the upper left corner, and ${a}$ to denote a scalar. The notation $|a|$ is the absolute value of $a$, $||\boldsymbol{A}||$ is the magnitude of $\boldsymbol{A}$ and determinant is shown by $\text{det}(\boldsymbol{A})$. Moreover, $\boldsymbol{A}^T$, $\boldsymbol{A}^H$ and $\boldsymbol{A}^*$ are the transpose, conjugate transpose and conjugate of $\boldsymbol{A}$, respectively. For a square matrix $\boldsymbol{A}$, $\boldsymbol{A}^{-1}$ represents its inverse. Also, $\boldsymbol{I}_N$ is the $N\times N$ identity matrix and $\lceil \cdot \rceil$ denotes the ceiling function.  A complex Gaussian random vector with mean $\boldsymbol{m}$ and covariance matrix $\boldsymbol{R}$ is shown by $\mathcal{C}\mathcal{N}(\boldsymbol{m},\boldsymbol{R})$, and $\text{E}[\boldsymbol{a}]$ and $\text{Cov}[\boldsymbol{a}]$ denote the expected value and covariance of ${\boldsymbol{a}}$, respectively.

\section{System Model}

\begin{figure}[t!]
\centering
\includegraphics[width=3.3in]{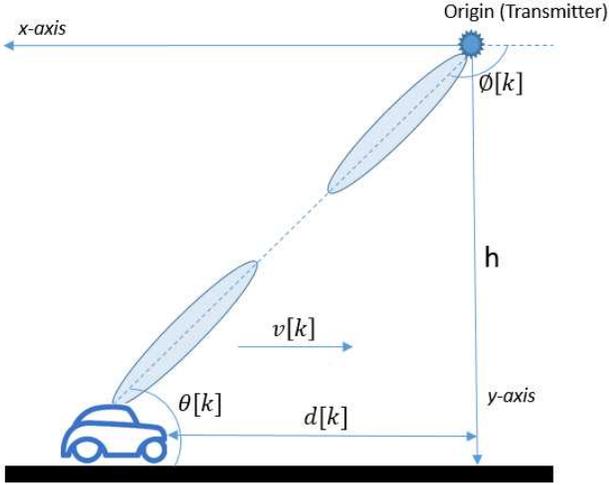}
\caption{Schematic of the physical model.}
\label{f1}
\end{figure}

Throughout the paper, we illustrate our proposed beam tracking approach for mmWave communications based on the schematic presented in Fig.~\ref{f1}. As can be seen in the figure, a base station is installed on an overpass with a height of $h$. On the receiver side, antennas are mounted on top of a vehicle moving with the speed of $v[k]$. As the index suggests, the speed of the car is considered to be variable at different transmission blocks.  

Moreover, the current location of the vehicle is specified based on its horizontal distance from the overpass denoted by $d[k]$ at the transmission block $k$. Also, the transmission and receiving angles of the vehicle are denoted by $\phi[k]$ and $\theta[k]$, respectively. Note that the angles are considered to be positive when they are between the positive $x$-axis and the line connecting the receiver to the transmitter.

\subsection{Transmission scheme}

Consider a mmWave MIMO system consisting of a transmitter with $N_t$ antennas and a receiver with $N_r$ antennas. The communication between the transmitter and receiver is broken down into $(K+1)$ discrete time blocks ($k=0,...,K$), and the duration of each block is represented by $\Delta t$. To simplify the notation, we set the value of $\Delta t$ to one. Therefore, the $i$th time block starts at the time $i$. 

Data transmission starts with the channel estimation in the first transmission block ($k=0$). The remaining blocks are dedicated to beam tracking until a certain error threshold is reached, and the re-estimation of channel is required. The channel estimation process is considered to be reasonably accurate \cite{rh,zhang}. Such accuracy can be achieved using several channel estimation techniques proposed for mmWave vehicular communications, or even the exhaustive search approach for beam tracking~\cite{gao2018dynamic
,li2018mobilize}. Our proposed approach in this paper only requires a single pilot transmission in the following blocks after the channel estimation ($k=1,..., K$) to track the estimated beam. It is particularly desirable to increase the tracking duration, as it leads to more vacancy for data transmission, and consequently, higher data transmission capacity.

The AoA and AoD of a single path between the transmitter and receiver at the $k$th transmission block are denoted by $\theta[k]$ and $\phi[k]$, respectively. Assuming uniform linear array (ULA) at both ends of the transmission, the receive and transmit array response vectors can be written as
\begin{align}\label{e25}
\nonumber\boldsymbol{a}_r&(\theta[k]) \\
&= \frac{1}{\sqrt{N_r}}[1,e^{-j\frac{2\pi}{\lambda}d\cos\theta[k]},...,e^{-j(N_r-1)\frac{2\pi}{\lambda}d\cos\theta[k]}]^T,\\
\nonumber\boldsymbol{a}_t&(\phi[k]) \\&= \frac{1}{\sqrt{N_t}}[1,e^{-j\frac{2\pi}{\lambda}d\cos\phi[k]},...,e^{-j(N_t-1)\frac{2\pi}{\lambda}d\cos\phi[k]}]^T,\label{e26}
\end{align}
where $N$, $d$, $\lambda$ denote the number of antennas, the antenna spacing, and the carrier wavelength, respectively. We have considered the model in two dimensions; however, the extension to three dimensions is straightforward according to \cite{el2014spatially}. 

Let $\overline{\theta}[k]$ and $\overline{\phi}[k]$ denote the pointing direction of the receiver combiner and the transmit beamformer , respectively, then the beamforming vectors can be written as
\begin{align} \label{e5}
\nonumber\boldsymbol{w}&(\overline{\theta}[k]) \\
&= \frac{1}{\sqrt{N_r}}[1,e^{-j\frac{2\pi}{\lambda}d\cos\overline{\theta}[k]},...,e^{-j(N_r-1)\frac{2\pi}{\lambda}d\cos\overline{\theta}[k]}]^T,\\
\nonumber\boldsymbol{f}&(\overline{\phi}[k]) \\&= \frac{1}{\sqrt{N_t}}[1,e^{-j\frac{2\pi}{\lambda}d\cos\overline{\phi}[k]},...,e^{-j(N_t-1)\frac{2\pi}{\lambda}d\cos\overline{\phi}[k]}]^T.
\label{e6}
\end{align}
The error in tracking the beam is the difference between the angles of beamformers and array response vectors at both ends of the transmission. Therefore, the objective is to maintain the difference between the angles as low as possible to track the beam for a longer period of time.  

\subsection{Channel model}
At the time $k$, the time-varying channel is shown as
\begin{equation} \label{e1}
\boldsymbol{H}[k] = \sum\limits_{l=1}^{L}\alpha_l[k]\boldsymbol{a}_r(\theta_{l}[k])\boldsymbol{a}_t^H(\phi_{l}[k]),
\end{equation}
where the index $l$ is the indicator of the $l$th path between the transmitter and receiver, and the variable $\alpha_l[k]$ shows the channel coefficient. The scattering in the mmWave spectrum has been shown to induce attenuation of over 20dB \cite{r11}. Hence, the line of sight beam is considered for the purpose of channel estimation and tracking \cite{shaham2018raf,shaham2018fast,booth2019multi,zhang2019fast}. This assumption is reasonable due to the recent measurements that showed  sparsity of mmWave communications \cite{r11,r12,p1}. Such sparsity indicates that the beams are separated in this spectrum, and it is highly likely that one path lies in the main beam direction with other paths falling into sidelobes \cite{rh}.

At the $k$th time block, after the transmission of pilot $s$ with unity power ($\|s\|_2 = 1$), the observed signal is given by
\begin{equation} \label{observation}
\begin{split}
z[k] &= \alpha[k]\boldsymbol{w}^H(\overline{\theta}[k])\boldsymbol{a_r}(\theta[k])\boldsymbol{a_t}^H(\phi[k])\boldsymbol{f}(\overline{\phi}[k])s + n[k],\\
\end{split}
\end{equation}
where $n[k]$ denotes the aggregated noise on the signal, and estimated as Gaussian process noise ($n[k]\thicksim \mathcal{C}\mathcal{N}(0,\sigma_n^2)$.


\section{Proposed beam tracking algorithm}\label{proposed approach}

The EKF algorithm is the extended version of the Kalman Filter that can be applied to nonlinear system models. The previous attempts on adopting the EKF algorithm for beam tracking in mmWave communications were predicated on using the AoA and AoD as state variables~\cite{rh,zhang}. As discussed earlier in the paper, there are several reasons why the current system model is not suitable for mmWave vehicular communications, such as non-linearity in the state evolution model, non-additive noise, and not being able to consider the kinematic characteristics of vehicular communications. These issues can be elaborated by examining the evolution of angles for a moving vehicle. If the vehicle moves from the transmit angle $\theta_{1}$ to $\theta_{2}$, the change in the transmit angle, and similarly, for the receiving angle is given by 
\begin{equation} \label{e4}
\theta_{2}-\theta_{1} = -\arctan\left(\frac{h}{(v_1 +w_1)\Delta t \cos^2(\theta_{1})} - \tan(\theta_{1})\right),
\end{equation}
where $v_1$ is the velocity of the vehicle at the first location and $w_1$ is the Gaussian noise\footnote{The derivation is excluded for the sake of brevity and can be found in the extended version of the paper in \cite{shaham2018fast}.}. It can be seen that (\ref{e4}) is non-linear with respect to the angle, as well as the process noise, which happens due to the non-additive nature of change in the velocity. If we are to use the angles as state variables, the calculation of the Jacobians for the EKF algorithm becomes highly complex and impractical. Hence, we propose to use the position, velocity, and complex channel gain as the state variables which give a linear state model and additive process noise for vehicular communications. The proposed state evolution model and observation equation are explained in the following, and the application of the EKF algorithm for beam tracking is illustrated.

\subsection{State Evolution Model}
Based on our proposed state variables, the corresponding state vector can be written as
\begin{equation} \label{e8}
\boldsymbol{x}[k] = [d[k],v[k],\alpha_R[k],\alpha_I[k]]^T,
\end{equation}
where $v[k]$ and $d[k]$ denote the velocity and position of the vehicle at the $k$th communication block, respectively. The Gaussian coefficient is split into a real part and an imaginary part, i.e., $\alpha[k] = \alpha_R[k] + j\alpha_I[k]$. Separating the real and imaginary parts of the coefficient helps to have all numbers in the state vector as real values. Both $\alpha_R[k]$ and $\alpha_I[k]$ follow the first-order Gauss-Markov model\cite{rh} expressed by
\begin{align} \label{e9}
\alpha_R[k+1] &=\rho \alpha_R[k] + \xi_1[k],\\
\alpha_I[k+1] &=\rho \alpha_I[k] + \xi_2[k],
\end{align}
where $\rho$ is the correlation coefficient, $\xi_1[k]$, $\xi_2[k] \thicksim \mathcal{N}\left(0,\dfrac{1-\rho^2}{2}\right)$, and  $\xi_1[-1]$, $\xi_2[-1] \thicksim \mathcal{N}\left(0,\dfrac{1}{2}\right)$. The evolution of the position and velocity of the vehicle in the next transmission block can be written as
\begin{align} \label{e11}
d[k+1] &= d[k] + v[k]\Delta t + w[k]\Delta t\\
v[k+1] &= v[k] + w[k],
\end{align}
with $w[k]$ denote the process noise representing the change in the speed of the vehicle. The value of $w[k]$ is assumed to follow Gaussian distribution $w[k]\thicksim \mathcal{N}(0,\sigma_w^2)$. In summary, the state evolution equation can be written as
\begin{equation} \label{e12}
\boldsymbol{x}[k+1] =\boldsymbol{A} \boldsymbol{x}[k] + \boldsymbol{u}[k],
\end{equation}
where
\begin{align} \label{e13}
\boldsymbol{A} =
\begin{bmatrix}
1\; \Delta t\;  0\;\;  0\\
0 \;\; 1 \;\; 0\;\; 0\\
0\;\;  0 \;\; 1\;\;  0\\
0\;\;  0 \;\; 0 \;\; 1
\end{bmatrix},
\end{align}
and $\boldsymbol{u}[k]\thicksim \mathcal{N}(0,\boldsymbol{\Sigma_u})$ with
\begin{align} \label{e14}
\boldsymbol{\Sigma_u} = diag(\,[\,(\Delta t\sigma_w)^2, (\sigma_w)^2, 1 - \rho^2, 1 - \rho^2\,]\,).
\end{align}

\subsection{Observation Expression}
To complete the scheme for the implementation of the EKF algorithm, we derive the measurement function based on the state variables. To start with, by substituting (\ref{e25}-\ref{e6}) in the observation equation in (\ref{observation}), we have


\begin{align} \label{e35}
\nonumber &z[k] = \frac{\alpha[k]s}{N_tN_r}\left(\sum\limits_{p=0}^{N_r-1}e^{-j\frac{2\pi}{\lambda}dp(\cos\theta[k]-\cos\overline{\theta}[k])}\right)\\\nonumber &\times\left(\sum\limits_{q=0}^{N_t-1}e^{j\frac{2\pi}{\lambda}dq(\cos\phi[k]-\cos\overline{\phi}[k])}\right)+ n[k]\\
&=\frac{\alpha[k]}{N_tN_r} \sum\limits_{q=0}^{N_t-1} \sum\limits_{p=0}^{N_r-1}e^{\frac{j2\pi d}{\lambda}(-p\cos\theta[k]+q\cos\phi[k]+b_{pq})}+ n[k],
\end{align}
 where
\begin{equation} \label{e28}
b_{pq} = p\cos\overline{\theta}[k]- q\cos\overline{\phi}[k].
\end{equation}
Based on Fig. \ref{f1}, the AoA and AoD of the system can be measured as
\begin{equation} \label{e15}
\theta[k] = \mathrm{atan2}\left(\frac{-h}{-d[k]}\right) = \mathrm{atan2}\left(\frac{-h}{-(d[k-1] + v[k-1]\Delta t)}\right)\\
\end{equation}
\begin{equation} \label{e16}
\phi[k] = \mathrm{atan2}\left(\frac{h}{d[k]}\right) = \mathrm{atan2}\left(\frac{h}{d[k-1] + v[k-1]\Delta t}\right).
\end{equation}
where $\mathrm{atan2}$ function returns the four-quadrant inverse tangent. The cosine of the angles are calculated as\footnote{$\cos(atan2(\dfrac{y}{x})) = \dfrac{y}{\sqrt{x^2+y^2}}$.}
\begin{align}\label{e29}
cos(\theta[k])&= \frac{-(d[k-1] + v[k-1]\Delta t)}{\sqrt{h^2+(d[k-1] + v[k-1]\Delta t)^2}}\\
cos(\phi[k])&= \frac{(d[k-1] + v[k-1]\Delta t)}{\sqrt{h^2+(d[k-1] + v[k-1]\Delta t)^2}}. \label{e30}
\end{align}
Substituting equations (\ref{e29}), (\ref{e30}) into (\ref{e35}), we have the observation equation in terms of the state variables as
\begin{align}  \label{e40}
\nonumber z&[k] =\frac{\alpha[k]s}{N_tN_r} \\ &r\times\sum\limits_{q=0}^{N_t-1} \sum\limits_{p=0}^{N_r-1}e^{\frac{j2\pi d}{\lambda}\left(\frac{(p+q)(d[k-1] + v[k-1]\Delta t)}{\sqrt{h^2+(d[k-1] + v[k-1]\Delta t)^2}}+b_{pq}\right)}+ n[k].
\end{align}
The first element of the above equation is a function of $\boldsymbol{x}[k]$. Therefore, (\ref{e40}) can be shown as

\begin{equation}
    z[k]= g(\boldsymbol{x}[k])+n[k].
\end{equation}

\subsection{EKF Based Beam Tracking}
Having developed the new state evolution model and the observation expression, the EKF algorithm can be applied for beam tracking in mmWave vehicular communications. The goal of the EKF algorithm is to match $\overline{\theta}[k]$ and $\overline{\phi}[k]$ to $\theta[k]$ and $\phi[k]$, respectively, as the vehicle moves. 

The EKF recursion algorithm~\cite{ekf} can be written as 
\begin{align}
&\text{Predicted state estimate:} \nonumber \\
&\widehat{\boldsymbol{x}}[k+1|k] = \boldsymbol{A}\widehat{\boldsymbol{x}}[k|k]\label{e41}\\
&\text{Predicted covariance estimate:} \nonumber \\
&\boldsymbol{P}[k+1|k] = \boldsymbol{A}\boldsymbol{P}[k|k] \boldsymbol{A}^T + \boldsymbol{\Sigma_u}\label{e42}\\
&\text{Innovation covariance:} \nonumber \\
&\boldsymbol{S}[k+1] = (\boldsymbol{H}[k+1] \boldsymbol{P}[k+1|k] \boldsymbol{H}[k+1]^T + \sigma_n^2)^{-1}\label{e43}\\
&\text{Kalman gain:} \nonumber \\
&\boldsymbol{K}[k+1] = \boldsymbol{P}[k+1|k] \boldsymbol{H}[k+1]^T \boldsymbol{S}[k+1]\label{e45}
\end{align}
$\text{\ \ Updated state estimate:}$
\begin{align} \label{e46}
\begin{split}
\widehat{\boldsymbol{x}}[k+1|k+1] &= \widehat{\boldsymbol{x}}[k+1|k]\\ &+ \boldsymbol{K}[k+1](z[k+1] - g(\widehat{\boldsymbol{x}}[k+1|k]))
\end{split}
\end{align}
$\text{\ \ Updated covariance estimate:}$
\begin{align} \label{e47}
&\boldsymbol{P}[k+1|k+1] = (\boldsymbol{I}-\boldsymbol{K}[k+1]\boldsymbol{H}[k+1]) \boldsymbol{P}[k+1|k],
\end{align}
where $\boldsymbol{H}[k+1]$ is the observation transition matrix defined by the following Jacobian matrix
\begin{equation} \label{e22}
\boldsymbol{H}[k+1] = \left.\frac{\partial g}{\partial\boldsymbol{x}}\right|_{\widehat{\boldsymbol{x}}[k+1|k]}.
\end{equation}
The closed-form expression for the partial derivative with respect to the position of the vehicle is derived in (\ref{e33}), 
\begin{figure*}[!t] 
\normalsize
\begin{align}
\frac{\partial z}{\partial d[k]} = \sum\limits_{q=0}^{N_t-1} \sum\limits_{p=0}^{N_r-1} \frac{\frac{\alpha[k]s}{N_tN_r}j\frac{2\pi}{\lambda}dh^2(p+q)}{\sqrt{(h^2+(d[k-1] + v[k-1]\Delta t)^2)^3}}  e^{\frac{j\frac{2\pi}{\lambda}d(b_{pq}\sqrt{h^2+(d[k-1] + v[k-1]\Delta t)^2} + (p+q)(d[k-1] + v[k-1]\Delta t)  )  }{\sqrt{h^2+(d[k-1] + v[k-1]\Delta t)^2}}} \label{e33}
\end{align}
\hrulefill
\end{figure*}
and the closed form expression for the partial derivative with respect the velocity can be calculated by
\begin{equation}\label{e36}
\frac{\partial z}{\partial v[k]} = \frac{\partial z}{\partial d[k]} \times \Delta t.
\end{equation}

The calculation of partial derivatives for the channel coefficient is straightforward, which corresponds to equation (\ref{e40}), excluding noise and channel coefficient. Note that the calculation of the Jacobian matrix in the current state depends on the Jacobians in the previous two states. Therefore, we set the initial states of the matrix to $\widehat{\boldsymbol{x}}_0$. Hence, we have
\begin{align*} \label{e20}
\widehat{\boldsymbol{x}}[0|0] &= \widehat{\boldsymbol{x}}[-1|-1] = \widehat{\boldsymbol{x}}_0,\\
\boldsymbol{P}[1|0] &= \boldsymbol{\Sigma_u}.
\end{align*}

The real-world interpretation of the initial state is the initial state of the vehicle, which is known based on the channel estimation. Furthermore, in order to deal with complex numbers in the implementation of EKF, $z[k]$ and $\boldsymbol{H}[k]$ are substituted by $\widetilde{z}[k] = [R(z[k]), I(z[k])]^T$ and $\widetilde{\boldsymbol{H}}_[k] = [R(\boldsymbol{H}[k]), I(\boldsymbol{H}[k])]^T$ in the equations (\ref{e41}-\ref{e47}).

\section{Numerical Results}

In this section, we investigate the performance of the proposed scheme for beam tracking based on the EKF algorithm in mmWave vehicular communications. First, the simulation setup is explained, followed by the performance analysis of the SNR and velocity. And ultimately, we compare our scheme to the previous work in~\cite{rh}. 

\subsection{Simulations Setup}

In our simulations, we assume that the transceivers are equipped with 16 antennas spaced by $\lambda/2$, and the initial value of AoD and AoA are set to $-135$ and $45$ degrees. Moreover, the values considered for the system variables are $\rho=0.995$, $\Delta t = 0.001$s, $h=3$m, and the initial speed of the vehicle is set to $60$ km/h. The received SNR and velocity are two significant factors affecting the performance of beam tracking, which will be investigated in the following.


\subsection{Effect of SNR on the estimation accuracy}
\begin{figure}[t!]
\centering
\includegraphics[width=\columnwidth]{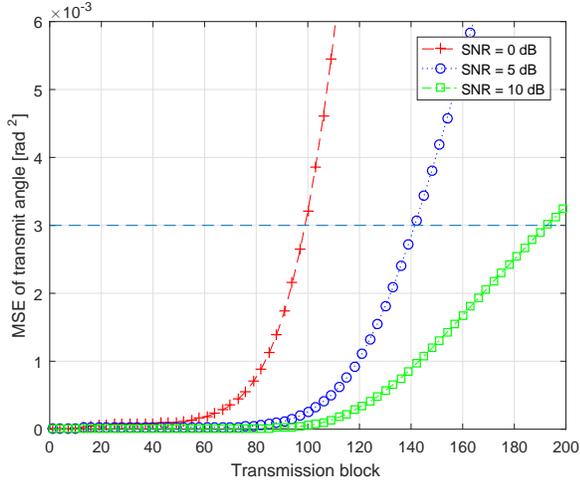}
\caption{Mean square error of the transmit angle versus the transmission block for different SNRs.}
\label{s2}
\end{figure}

The mean square error (MSE) performance of the EKF tracking algorithm for different SNR values is shown in Fig.~\ref{s2}. Each point in the graph was obtained after 3000 runs of the algorithm. Note that the performance is given for the MSE of the transmit angle, and similarly, it can be derived for the AoA. In our simulations, the valid tracking threshold is chosen to be $\sqrt{E[|\phi[k]-\overline{\phi}[k]|^2 ]} = BW/2$ \cite{rh}, where $BW$ denotes the beamwidth. The tracking of a beam is said to be lost if the value of MSE exceeds the threshold shown by a horizontal line on the figure. As can be seen in Fig. \ref{s2}, the curves for SNRs$=0,\,5,\,10$dB cross the threshold in 99, 142 and 193 transmission blocks, respectively, where each transmission block corresponds to 1$ms$. As expected, increasing the value of SNR results in tracking for a longer period of time.

\subsection{Impact of Velocity on the Estimation Accuracy}
\begin{figure}[t!]
\centering
\includegraphics[width=\columnwidth]{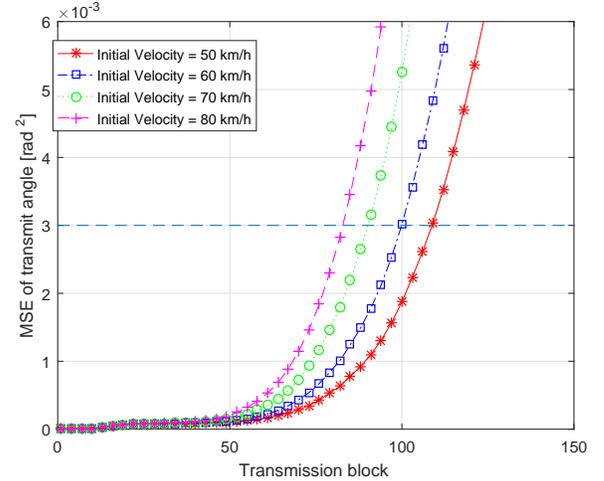}
\caption{Mean square error of the transmit angle versus the transmission block size at different initial speed of vehicle.}
\label{s3}
\end{figure}

Fig. \ref{s3} demonstrates the impact of the velocity of the vehicle on beam tracking. The graph is plotted for SNR=0dB, and the same variance of the velocity process noise ($\sigma_w=0.28 $). As can be seen in the figure, increasing the speed negatively affects the performance of beam tracking. This is mainly due to the fact that the duration of the transmission block is constant, and according to the kinematic formula, we have  $(position\,displacement) = (velocity) \times (block\,duration)$. Hence, for a higher velocity, we expect a larger displacement of the vehicle in between the transmission blocks. Therefore, it is reasonable that the beam tracking performance deteriorates as the initial speed increases. The summary of the valid tracking duration for various initial velocities is presented in Table \ref{t1}.  
\begin{table}[t]
\caption{Valid tracking duration for different initial speeds.}
\centering
\footnotesize
\begin{tabular}{|c | c | c | c | c |}
 \hline
 Initial velocity & 50 km/h & 60 km/h & 70 km/h & 80 km/h\\
 \hline
 AoD (time slot) & 109 & 100 & 91 & 83\\
 \hline
\end{tabular}
\label{t1}
\end{table}

\subsection{Comparison}
We compare our proposed scheme and state variables for the EKF algorithm with the previously developed approach in \cite{rh}. The authors in \cite{rh} use the AoD, AoA, and the channel coefficient as their state variables to track the beam. However, the assumption behind the approach is that the angles evolve based on a process noise with zero mean and variance $(\dfrac{0.5}{180}\pi)^2$. Moreover, the noise was considered to be additive, and the state evolution equations were deemed to be linear. As explained in Section \ref{proposed approach}, such characterization cause several complications in vehicular settings. If the angles are used as state variables of the system model to track vehicles, the state evolution of the system is no longer linear, and the noise is undoubtedly not additive.  Such nonlinearity leads to high complexity in the calculation of Jacobians for beam tracking, making it an impractical approach for mmWave vehicular communications. 

\begin{figure}[t!]
\centering
\includegraphics[width=3.46in, height=2.4in]{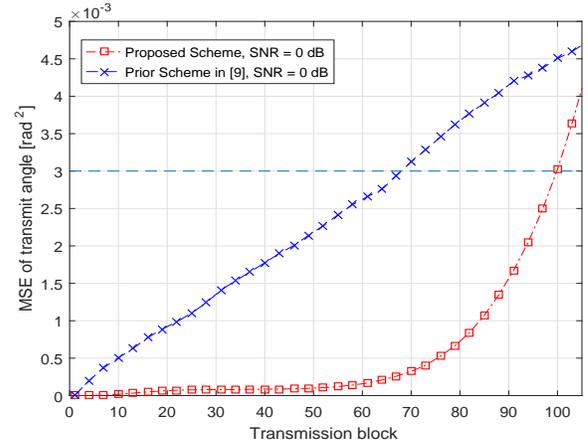}
\caption{Mean square error of the transmit angle versus teh transmission block size at SNR=0dB for the proposed scheme and the approach developed in  \cite{rh}.}
\label{s4}
\end{figure}

Fig.~\ref{s4} shows the comparison between our proposed approach and the scheme proposed in \cite{rh}. Our suggested scheme results in two significant improvements: (I) The valid tracking performance of the beam is improved by $49\%$, particularly in high SNRs. (II) Our proposed scheme considers the realistic characteristics of the vehicular communication setting, such as the velocity, the variation in the speed of the vehicle, and the transmission block duration.



\section{conclusion}

In this paper, we proposed a scheme for the implementation of beam tracking based on the EKF algorithm in mmWave vehicular communications. New state evolution and observation models were proposed based on the position, velocity, and channel coefficient. The proposed scheme generates linear state evolution, which significantly simplifies the implementation of the EKF beam tracking. The Jacobians were derived in closed-form expressions, which opens up an efficient way to run the EKF algorithm. The new scheme improves beam tracking performance in mmWave vehicular communications by considering the dynamics of vehicular communication systems.

\bibliographystyle{IEEEtran}
\bibliography{beamtracking_s}

\begin{thebibliography}{10}
\providecommand{\url}[1]{#1}
\csname url@samestyle\endcsname
\providecommand{\newblock}{\relax}
\providecommand{\bibinfo}[2]{#2}
\providecommand{\BIBentrySTDinterwordspacing}{\spaceskip=0pt\relax}
\providecommand{\BIBentryALTinterwordstretchfactor}{4}
\providecommand{\BIBentryALTinterwordspacing}{\spaceskip=\fontdimen2\font plus
\BIBentryALTinterwordstretchfactor\fontdimen3\font minus
  \fontdimen4\font\relax}
\providecommand{\BIBforeignlanguage}[2]{{%
\expandafter\ifx\csname l@#1\endcsname\relax
\typeout{** WARNING: IEEEtran.bst: No hyphenation pattern has been}%
\typeout{** loaded for the language `#1'. Using the pattern for}%
\typeout{** the default language instead.}%
\else
\language=\csname l@#1\endcsname
\fi
#2}}
\providecommand{\BIBdecl}{\relax}
\BIBdecl

\bibitem{fazliu2019mmwave}
Z.~L. Fazliu, F.~Malandrino, and C.-F. Chiasserini, ``mmwave in vehicular
  networks: Leveraging traffic signals for beam design,'' 2019.

\bibitem{kawser2019perspective}
M.~T. Kawser, M.~S. Fahad, S.~Ahmed, S.~S. Sajjad, and H.~A. Rafi, ``The
  perspective of vehicle-to-everything (v2x) communication towards 5g,''
  \emph{IJCSNS}, vol.~19, no.~4, p. 146, 2019.

\bibitem{ghosh2019inclusive}
S.~Ghosh and D.~Sen, ``An inclusive survey on array antenna design for
  millimeter-wave communications,'' \emph{IEEE Access}, vol.~7, pp.
  83\,137--83\,161, 2019.

\bibitem{r2}
V.~Va, T.~Shimizu, G.~Bansal, R.~W. Heath~Jr \emph{et~al.}, ``Millimeter wave
  vehicular communications: A survey,'' \emph{Foundations and
  Trends{\textregistered} in Networking}, vol.~10, no.~1, pp. 1--113, 2016.

\bibitem{r3}
J.~Choi, V.~Va, N.~Gonzalez-Prelcic, R.~Daniels, C.~R. Bhat, and R.~W. Heath,
  ``Millimeter-wave vehicular communication to support massive automotive
  sensing,'' \emph{IEEE Communications Magazine}, vol.~54, no.~12, pp.
  160--167, December 2016.

\bibitem{midc}
K.~Hosoya, N.~Prasad, K.~Ramachandran, N.~Orihashi, S.~Kishimoto,
  S.~Rangarajan, and K.~Maruhashi, ``Multiple sector id capture (midc): A novel
  beamforming technique for 60-ghz band multi-gbps wlan/pan systems,''
  \emph{IEEE Transactions on Antennas and Propagation}, vol.~63, no.~1, pp.
  81--96, Jan 2015.

\bibitem{midc2}
Y.~Inoue, Y.~Kishiyama, Y.~Okumura, J.~Kepler, and M.~Cudak, ``Experimental
  evaluation of downlink transmission and beam tracking performance for 5g mmw
  radio access in indoor shielded environment,'' in \emph{2015 IEEE 26th Annual
  International Symposium on Personal, Indoor, and Mobile Radio Communications
  (PIMRC)}, Aug 2015, pp. 862--866.

\bibitem{zhang}
C.~Zhang, D.~Guo, and P.~Fan, ``Tracking angles of departure and arrival in a
  mobile millimeter wave channel,'' in \emph{2016 IEEE International Conference
  on Communications (ICC)}, May 2016, pp. 1--6.

\bibitem{rh}
V.~Va, H.~Vikalo, and R.~W. Heath, ``Beam tracking for mobile millimeter wave
  communication systems,'' in \emph{2016 IEEE Global Conference on Signal and
  Information Processing (GlobalSIP)}, Dec 2016, pp. 743--747.

\bibitem{gao2018dynamic}
M.~Gao, B.~Ai, Y.~Niu, Z.~Zhong, Y.~Liu, G.~Ma, Z.~Zhang, and D.~Li, ``Dynamic
  mmwave beam tracking for high speed railway communications,'' in \emph{2018
  IEEE Wireless Communications and Networking Conference Workshops
  (WCNCW)}.\hskip 1em plus 0.5em minus 0.4em\relax IEEE, 2018, pp. 278--283.

\bibitem{li2018mobilize}
J.~Li, Y.~Sun, L.~Xiao, S.~Zhou, and A.~Sabharwal, ``How to mobilize mmwave: A
  joint beam and channel tracking approach,'' in \emph{2018 IEEE International
  Conference on Acoustics, Speech and Signal Processing (ICASSP)}.\hskip 1em
  plus 0.5em minus 0.4em\relax IEEE, 2018, pp. 3624--3628.

\bibitem{el2014spatially}
O.~El~Ayach, S.~Rajagopal, S.~Abu-Surra, Z.~Pi, and R.~W. Heath, ``Spatially
  sparse precoding in millimeter wave mimo systems,'' \emph{IEEE transactions
  on wireless communications}, vol.~13, no.~3, pp. 1499--1513, 2014.

\bibitem{r11}
S.~Han, C.~l.~I, Z.~Xu, and C.~Rowell, ``Large-scale antenna systems with
  hybrid analog and digital beamforming for millimeter wave 5g,'' \emph{IEEE
  Communications Magazine}, vol.~53, no.~1, pp. 186--194, January 2015.

\bibitem{shaham2018raf}
S.~Shaham, M.~Kokshoorn, Z.~Lin, M.~Ding, and Y.~Wu, ``Raf: Robust adaptive
  multi-feedback channel estimation for millimeter wave mimo systems,'' in
  \emph{2018 IEEE Wireless Communications and Networking Conference
  (WCNC)}.\hskip 1em plus 0.5em minus 0.4em\relax IEEE, 2018, pp. 1--6.

\bibitem{shaham2018fast}
S.~Shaham, M.~Ding, M.~Kokshoorn, Z.~Lin, and X.~Yao, ``Fast channel estimation
  and beam tracking for millimeter wave vehicular communications,'' \emph{arXiv
  preprint arXiv:1806.00161}, 2018.

\bibitem{booth2019multi}
M.~B. Booth, V.~Suresh, N.~Michelusi, and D.~J. Love, ``Multi-armed bandit beam
  alignment and tracking for mobile millimeter wave communications,''
  \emph{IEEE Communications Letters}, 2019.

\bibitem{zhang2019fast}
D.~Zhang, A.~Li, M.~Shirvanimoghaddam, P.~Cheng, Y.~Li, and B.~Vucetic, ``Fast
  beam tracking for millimeter-wave systems under high mobility,'' in \emph{ICC
  2019-2019 IEEE International Conference on Communications (ICC)}.\hskip 1em
  plus 0.5em minus 0.4em\relax IEEE, 2019, pp. 1--6.

\bibitem{r12}
L.~Dai and X.~Gao, ``Priori-aided channel tracking for millimeter-wave
  beamspace massive mimo systems,'' in \emph{2016 URSI Asia-Pacific Radio
  Science Conference (URSI AP-RASC)}, Aug 2016, pp. 1493--1496.

\bibitem{p1}
R.~W. Heath, N.~González-Prelcic, S.~Rangan, W.~Roh, and A.~M. Sayeed, ``An
  overview of signal processing techniques for millimeter wave mimo systems,''
  \emph{IEEE Journal of Selected Topics in Signal Processing}, vol.~10, no.~3,
  pp. 436--453, April 2016.

\bibitem{ekf}
T.~Kailath, A.~H. Sayed, and B.~Hassibi, \emph{Linear estimation}.\hskip 1em
  plus 0.5em minus 0.4em\relax Prentice Hall Upper Saddle River, NJ, 2000,
  vol.~1.

\end{thebibliography}

\end{document}